\documentclass[reprint, showpacs, preprintnumbers, amsmath,amssymb, aps]{revtex4-1}

\usepackage[colorlinks=true,linkcolor=black, citecolor=black,
urlcolor=black]{hyperref}
\usepackage{adjustbox}

\usepackage{multirow,graphics}

\usepackage{amstext}
\usepackage{amssymb}
\usepackage{amsmath}
\usepackage{graphicx}
\usepackage{colonequals}
\usepackage{mathrsfs}

\usepackage[usenames,dvipsnames]{color}

\begin{document}

\newcommand{\abra}[2]{\ensuremath{\langle #1 #2\rangle}}
\newcommand{\sbra}[2]{\ensuremath{[ #1 #2]}}
\newcommand{\form}[2]{\ensuremath{\left\{ #1 #2\right\}}}
\newcommand{\bradot}[3]{\ensuremath{\langle #1|\, #2|#3]}}

\newcommand{\calN}{\mathcal{N}}\newcommand{\calM}{\mathcal{M}}
\newcommand{\calS}{\mathcal{S}}\newcommand{\calF}{\mathscr{F}}
\newcommand{\calZ}{\mathcal{Z}}\newcommand{\calC}{\mathcal{C}}
\newcommand{\calA}{\mathcal{A}}\newcommand{\calO}{\mathcal{O}}
\newcommand{\calOg}{\mathcal{O}}
\newcommand{\calOs}{\mathcal{O}'}

\newcommand{\Tr}{\text{Tr}}
\newcommand{\CC}{\mathbb{C}}
\newcommand{\RR}{\mathbb{R}}
\newcommand{\intd}{\int\mathrm{d}}
\newcommand{\dd}{\mathrm{d}}
\newcommand{\DD}{\mathrm{D}}
\newcommand{\AAA}{\mathcal{A}}
\newcommand{\WWW}{\mathcal{W}}
\newcommand{\ZZZ}{\mathcal{Z}}
\newcommand{\FFF}{\mathcal{F}}
\newcommand{\OOO}{\mathcal{O}}
\newcommand{\PPP}{\mathbf{F}}
\newcommand{\la}{\lambda}
\newcommand{\lat}{\tilde{\lambda}}
\newcommand{\si}{\sigma}
\newcommand{\sit}{\tilde{\sigma}}
\newcommand{\m}{\textbf{m}}\newcommand{\n}{\textbf{n}}
\newcommand{\fA}{\mathsf{A}}
\newcommand{\fP}{\mathsf{P}}\newcommand{\fQ}{\mathsf{Q}}
\newcommand{\fp}{\mathsf{p}}
\newcommand{\ftau}{\mathsf{\tau}}
\newcommand{\fq}{\mathsf{q}}
\newcommand{\fR}{\mathsf{R}}
\newcommand{\fw}{\mathsf{w}}
\newcommand{\iconst}{4\pi^2}
\newcommand{\constA}{-2\pi i}
\newcommand{\constB}{\frac{1}{2(2\pi i)^2}}

\newcommand{\beq}{\begin{equation}}
\newcommand{\eeq}{\end{equation}}
\newcommand{\eqndot}{\, .}
\newcommand{\eqncom}{\, ,}
\newcommand{\eqnsem}{\, ;}
\newcommand{\YM}{{\mathrm{\scriptscriptstyle YM}}}
\newcommand{\phaneq}{\phantom{{}=}}

\newcommand{\e}{\operatorname{e}}

\newcommand{\bra}[1]{\ensuremath{\left< #1\,\right|}}
\newcommand{\ket}[1]{\ensuremath{\left|\, #1\right>}}
\newcommand{\vac}[1]{\ensuremath{\left< \, #1\, \right>}}
\newcommand{\vev}[3]{\ensuremath{\big< #1\,\big| #2\,\big| #3\big>}}


\preprint{HU-Mathematik-16-05, 
 \ HU-EP-16/09,\ MITP/16-024}

\title{\texorpdfstring{Composite Operators in the Twistor Formulation of $\boldsymbol{\mathcal{N}=4}$ SYM Theory}{Composite Operators in the Twistor Formulation of N=4 SYM Theory}}

\author{Laura Koster$^{a}$}
\email{laurakoster@physik.hu-berlin.de}
\author{Vladimir Mitev$^{a,b}$}
\email{vmitev@uni-mainz.de}
\author{Matthias Staudacher$^a$}
\email{matthias@math.hu-berlin.de}
\author{Matthias Wilhelm$^{a,c}$}
\email{matthias.wilhelm@nbi.ku.dk}

\affiliation{
\(^{a}\)Institut f\"ur Mathematik und Institut f\"ur Physik, Humboldt-Universit\"at zu Berlin, 
IRIS Haus, Zum Gro{\ss}en Windkanal 6,  12489 Berlin, Germany
\\
\(^{b}\)PRISMA Cluster of Excellence, Institut f\"ur Physik, WA THEP, 
Johannes Gutenberg-Universit\"at Mainz, 
Staudingerweg 7, 55128 Mainz, Germany
\\
\(^{c}\)Niels Bohr Institute, Copenhagen University, 
Blegdamsvej 17, 2100 Copenhagen \O{}, Denmark  }

\begin{abstract}
We incorporate gauge-invariant local composite operators into the twistor-space formulation of $\calN=4$ Super Yang-Mills theory.
In this formulation, the interactions of the elementary fields are reorganized into infinitely many interaction vertices and we argue that the same applies to composite operators.
To test our definition of the local composite operators in twistor space, we compute several corresponding form factors, thereby also initiating the study of form factors using the position twistor-space framework. 
Throughout this letter, we use the composite operator built from two identical complex scalars as a pedagogical example; we treat the general case in a follow-up paper.
\end{abstract}


 \maketitle
\section{Introduction}

The study of the simplest interacting gauge theory in four dimensions, namely $\mathcal{N}=4$ Super Yang-Mills (SYM), has led to a plethora of important theoretical insights, such as holography (AdS/CFT), integrability in the planar limit, on-shell methods and more. Moreover, the theory can be formulated in twistor space \cite{Boels:2006ir}, which has been an efficient setting for computing on-shell quantities such as amplitudes \cite{ArkaniHamed:2009dn,Mason:2009qx,Bullimore:2010pj,Adamo:2011cb,Adamo:2011pv} and relating them to light-like Wilson loops \cite{CaronHuot:2010ek, Mason:2010yk,Belitsky:2011zm}. 
The action of $\calN=4$ SYM in twistor space is the sum of two parts $\calS_1+\calS_2$, with $\calS_1$, introduced in  \cite{Witten:2003nn}, describing the self-dual part and $\calS_2$, referred to as the interaction piece,
 immediately giving the maximally-helicity-violating (MHV) tree-level scattering amplitudes \cite{Boels:2007qn}. 
However, computations involving off-shell quantities, such as form factors or correlation functions of gauge-invariant local composite operators (composite operators), are less straightforward in the twistor formalism, though some progress has been made \cite{Koster:2014fva, Chicherin:2014uca}. To tackle off-shell objects, one needs a proper definition of composite operators in twistor space, which is the main subject of this letter 
\footnote{The problem addressed in this paper does not occur in \cite{Chicherin:2014uca} due to the use of the Lagrangian-insertion technique and it was explicitly deferred to this paper in \cite{Koster:2014fva}.}.

In the usual space-time formulation, the composite operator $\calOg$ consists of a single term that immediately determines its vertex. In contradistinction, we argue that the twistor-space representation of $\calOg$ (or rather its vertex) must contain infinitely many terms as it has to describe all interactions of $\calOg$ with elementary particles at minimal MHV degree.
In particular, all MHV tree-level form factors of the operator $\calOg$ have to be given immediately by the operator vertex; an elementary counting of the MHV degree shows that they cannot contain any twistor-space propagators and hence also no interaction vertices.

In this letter, we explicitly demonstrate this principle using the gauge-invariant local composite operator \begin{equation}\calOs=\frac{1}{2}\Tr[\phi_{ab}^2]\end{equation} built out of two identical complex scalars. 
In particular, we determine the correct twistor-space vertex for this operator.
An algorithm for generating the vertices for all operators using Wilson loops will be given in a follow-up paper.

The definition of a composite operator $\calOg$ in twistor space can be probed by computing its tree-level MHV form factor with external on-shell states $A_1, \ldots, A_n$ and comparing this to data from the literature. Tree-level MHV form factors are the simplest quantities that contain a composite operator and hence provide an ideal testing ground for our definition of composite operators in twistor space
\footnote{In \cite{Brandhuber:2011tv}, 
dual MHV rules, which can be rewritten in terms of twistor diagrams, were used to study form factors of $\calOs$. Whereas the twistors there correspond to dual momentum space, we are studying form factors via twistors in position space.}.
Letting $\fp_i$ be the momenta of the on-shell states, and $\fq$ the momentum of $\calOg$, the form factor is defined as the expectation value
\begin{multline}
\label{eq:formfactordefinition}
\calF_{\calOg}(1^{A_1},\ldots, n^{A_n};\fq)
\\= \int \frac{\dd^4x}{(2\pi)^4}\, \e^{-i\fq x}\vev{A_1(\fp_1)\cdots A_n(\fp_n)}{\calOg(x)}{0}\,.
\end{multline}
Form factors in $\mathcal{N}=4$ SYM are also interesting in their own right and have
 received increasing attention, both at weak coupling \cite{vanNeerven:1985ja,Brandhuber:2010ad,Bork:2010wf,Brandhuber:2011tv,
Bork:2011cj,Henn:2011by,Gehrmann:2011xn,Brandhuber:2012vm,Bork:2012tt,
Engelund:2012re,Johansson:2012zv,Boels:2012ew,Penante:2014sza,
Brandhuber:2014ica,Bork:2014eqa,Wilhelm:2014qua,Nandan:2014oga,Loebbert:2015ova,
Bork:2015fla,Frassek:2015rka,Boels:2015yna,Huang:2016bmv} and at strong coupling \cite{Alday:2007he,Maldacena:2010kp,Gao:2013dza}. 
In comparison to amplitudes, however, where all tree-level expression \cite{Drummond:2008cr} as well as the unregularized integrand of all loop-level expressions \cite{ArkaniHamed:2010kv} have been found, much less is known for form factors; see \cite{Wilhelm:2016izi} for the state of the art. 
In particular, the form factors of the operator $\calOs$ are also phenomenologically interesting, as they are related to the Higgs-to-gluons amplitude in QCD; cf.\ \cite{Brandhuber:2012vm}.

\section{A first attempt at composite operators\,\ldots}
\label{sec:firstattempt}

We refer to \cite{Adamo:2013cra} for an introduction to the supertwistor methods that we shall use here. We write supertwistors $\calZ\in \mathbb{CP}^{3|4}$ as $\calZ =  (\la_{\alpha},\mu^{\dot\alpha},\chi^a)$, where the $\chi^a$ are fermionic, $\alpha, \dot{\alpha}\in \{1,2\}$ and  $a\in\{1,2,3,4\}$. Supertwistor space is naturally related to chiral Minkowski superspace $\mathbb{M}^{4|8}$, which is obtained by appending eight Gra\ss mann variables $\theta^{\alpha a}$ to each point $x^{\alpha\dot\alpha}$ in Minkowski space.
Each point $(x,\theta)$ in $\mathbb{M}^{4|8}$ corresponds to a unique projective line in supertwistor space  given by the set of supertwistors 
\begin{equation}
\label{eq:definitioncalZonaline}
\calZ=(\la_{\alpha},ix^{\alpha\dot\alpha}\la_{\alpha},i\theta^{a\alpha}\la_{\alpha})\,,\quad \lambda\in \mathbb{CP}^1\,.
\end{equation}
For brevity, we denote a line in supertwistor space by $x$ instead of $(x,\theta)$ and we denote by $\calZ_x(\lambda)$  the supertwistor \eqref{eq:definitioncalZonaline} on the line $x$ given by the spinor $\lambda$.

In order to obtain a field $\Phi$ in space-time, the standard prescription is to Penrose transform \cite{Penrose:1969ae, Penrose:1978qb} a field in twistor space $\tilde{\Phi}$, i.e.\ integrating the twistor-space field $\tilde{\Phi}$ over the line in $\mathbb{CP}^{3|4}$ corresponding to $(x,\theta)$ as
\begin{equation}
\label{eq:Penrose transform}
\Phi(x)=\int_{\mathbb{CP}^1}\DD\la\, \tilde{\Phi}(\calZ_{x}(\lambda))\,,
\end{equation} 
where $\DD\la=\frac{\abra{\lambda}{\dd\lambda}}{2\pi i}$. 
The angle bracket is defined as $\abra{\lambda}{\lambda'}= \lambda^{\alpha}\lambda'_{\alpha}$ with $\epsilon^{\alpha\beta}\lambda_{\alpha}=\lambda^{\beta}$ and $\epsilon^{12}=1$.  
For future reference, we note that integrals over the spinors $\lambda$ are always taken over the projective line $\mathbb{CP}^1$, so that we can omit this from the integral sign.

The  twistor action \cite{Boels:2006ir} is written using a single connection superfield $\calA$  introduced in \cite{Nair:1988bq}. This superfield combines the on-shell degrees of freedom of $\calN=4$ SYM -- the two helicity $\pm 1$ gluons $g^{\pm}$, the four helicity $\frac{1}{2}$ fermions $\bar{\psi}_a$ and their antiparticles $\psi^a$ and the six scalars $\phi_{ab}$ -- as
\begin{multline}
\label{eq:expansionAAA}
\AAA(\mathcal{Z})=g^+ +\chi^a\bar{\psi}_a+\frac{1}{2}\chi^a\chi^b\phi_{ab}\\+\frac{1}{3!}\chi^a\chi^b\chi^c\psi^d\epsilon_{abcd}+\chi^1\chi^2\chi^3\chi^4 g^-\eqncom
\end{multline}
where the components fields $g^\pm,\ldots$ do not depend on the Gra\ss mann variables $\chi$.
According to \eqref{eq:Penrose transform}, a natural first attempt~\cite{Adamo:2011cd, Adamo:2011dq} for the (gauge-covariant) scalar field $\phi_{ab}(x)$ is given by a Penrose transform
\begin{equation}
\label{eq:naturalfirstattempt}
\phi_{ab}(x)\stackrel{?}=\int \DD\la \, h_x^{-1}(\la)\frac{\partial^2\calA(\lambda)}{\partial\chi^a\partial\chi^b}h_x(\la)_{\big{|}\theta=0}\,,
\end{equation} 
where $\calA(\lambda)\equiv \calA(\calZ_x(\lambda))$. In \eqref{eq:naturalfirstattempt}, we set $\theta=0$, implying $\chi^a=i\theta^{\alpha a}\lambda_{\alpha}=0$, after taking the derivatives, because we are only interested in the $\phi_{ab}$ component of \eqref{eq:expansionAAA}.
In addition, we have introduced $h_x(\la)$ -- the frame on $x$ that trivializes the connection $\calA$ along the line $x$ and thus ensures gauge invariance when taking traces of products of these fields \cite{Adamo:2011dq}.

Therefore, using the ansatz \eqref{eq:naturalfirstattempt}, the operator $\calOs$ built out of two scalars would read
\begin{multline}
\label{eq:operator}
\calOs(x)=
\frac{1}{2}\Tr[\phi_{ab}^2](x)\stackrel{?}= \frac{1}{2}\int \DD\la \DD{\la'}\,\Tr\Big[\frac{\partial^2\calA(\la)}{\partial\chi^a\partial\chi^b}\\\times U_x(\la,\la')\frac{\partial^2\calA(\la')}{\partial\chi'^a\partial\chi'^b}U_x(\la',\la)\Big]_{\big{|}\theta=0}\,,
\end{multline}
where $U_x(\la,\la')=h_x(\la)h_x(\la')^{-1}$ is the parallel propagator for the connection $\AAA$. It can be expanded as 
\begin{multline}
\label{eq:frameUdefinitionpart1}
 U_{x}(\la,\la')\equiv 
  U_{x}(\calZ_x(\la),\calZ_x(\la'))\\
=1+\sum_{m=1}^\infty\int \frac{\abra{\lambda}{\lambda'}\DD\tilde{\la}_1\cdots \DD\tilde{\la}_m\AAA(\tilde{\la}_1) \cdots \AAA(\tilde{\la}_m)}{\abra{\lambda}{\tilde{\la}_1}\abra{\tilde{\la}_1}{\tilde{\la}_2}\cdots \abra{\tilde{\la}_m}{\lambda'}}\eqndot
\end{multline}

In order to obtain form factors \eqref{eq:formfactordefinition}, which are naturally expressed  in momentum space, we insert external on-shell momentum states \cite{Adamo:2011cb} of (super-)momentum $\fP=(\fp_{\alpha\dot\alpha},\eta_a)= (p_{\alpha},\bar{p}_{\dot \alpha},\eta_a)$:
\begin{equation}
 \label{eq:definitiononshellmomentumeigenstates}
\calA_{\fP}(\calZ)=2\pi i \int_{\mathbb{C}}\frac{\dd s}{s}\e^{s(\mu^{\dot{\alpha}}\bar{p}_{\dot{\alpha}}+\chi^a\eta_{a})}
\bar{\delta}^2(s\lambda-p)\eqncom
\end{equation}
where $\bar{\delta}^2(\lambda)= \bar{\delta}^1(\lambda_1)\bar{\delta}^1(\lambda_2)$ with the $\bar{\delta}^1(z)$  denoting the $\delta$ function on the complex plane.
Let us compute $\calF_{\calOs}(1^{+},\ldots, i^{\phi_{ab}},\ldots, j^{\phi_{ab}},\ldots, n^{+};\fq)$: the (color-ordered) form factor of $\calOs$ with $n$ external particles, two of which are scalars $\phi_{ab}$ at positions $i$ and $j$, while the remaining ones are positive-helicity gluons. Here, only the terms in \eqref{eq:frameUdefinitionpart1} with the appropriate number of $\calA$'s contribute, namely those with $j-i-1$ from one $U_x$ and $n+i-j-1$ from the other. Inserting the on-shell states \eqref{eq:definitiononshellmomentumeigenstates} into \eqref{eq:frameUdefinitionpart1} as well as directly into  \eqref{eq:operator} and then integrating over $s$ and the corresponding $\la$ effectively cancels $s$ and replaces $\la_{\alpha}\rightarrow p_{\alpha}$, $\mu^{\dot{\alpha}}\rightarrow ix^{\alpha \dot{\alpha}}p_{\alpha}$ and $\chi^a\rightarrow i\theta^{\alpha a }p_{\alpha}$ due to the $\bar{\delta}^2$ function and the parametrization \eqref{eq:definitioncalZonaline}. Selecting the coefficient of the corresponding $\eta$'s
and Fourier transforming in $x$ as $\int \frac{\dd^4x}{(2\pi)^4} \e^{-i\fq x}$ yields the desired form factor 
\begin{multline}
\label{eq:formfactor2phi}
\calF_{\calOs}(1^{+},\ldots, i^{\phi_{ab}},\ldots, j^{\phi_{ab}},\ldots, n^{+};\fq)\\
=-\frac{\abra{i}{j}^2\delta^4(\fq-\sum_{k=1}^n\fp_k)}{\abra{1}{2}\cdots \abra{n}{1}}\,,
\end{multline}
which perfectly agrees with the result of \cite{Brandhuber:2010ad}. We would like to comment that the parallel propagators $U_x(\la,\la')$ in \eqref{eq:operator}, introduced in order to ensure gauge invariance, are responsible for inserting infinitely many vertices in the description of the composite operators related to their MHV coupling to positive-helicity gluons.

\section{\ldots\,and where it fails}

While the successful computation of the form factor \eqref{eq:formfactor2phi} is encouraging, the ansatz \eqref{eq:operator} fails once one considers external states involving fermions. Specifically, let us take the MHV form factor
\beq
\label{eq:formactorpsisquared}
\calF_{\calOs}(1^{\bar{\psi}_a},2^{\bar{\psi}_b},3^{\phi_{ab}};\fq)
=\frac{\delta^4(\fq-\sum_{k=1}^3\fp_k)}{\abra{1}{2}}\,,
\eeq
which was first calculated in \cite{Brandhuber:2011tv}.  However, using the twistor-space machinery and \eqref{eq:operator}, we would obtain zero. This can be seen as follows. On the one hand, the contribution to the form factor \eqref{eq:formactorpsisquared} has to come solely from the  operator $\calOs$ itself, since the inclusion of an interaction vertex from $\calS_2$ connected by a propagator to $\calOs$ would increase the MHV degree. On the other hand, every form factor obtained from \eqref{eq:operator} will necessarily contain two on-shell scalars due to the $\frac{\partial^2\calA}{\partial \chi^a\partial \chi^b}$ terms, while  \eqref{eq:formactorpsisquared} contains only one.

We conclude that it is necessary to add extra terms encoding the contribution to the form factor \eqref{eq:formactorpsisquared}  directly into the twistor-space expression of the operator $\calOs$.

\section{Our proposal}
\label{sec:ourproposal}

\begin{figure}[htbp]
 \centering
  \includegraphics[height=2.5cm]{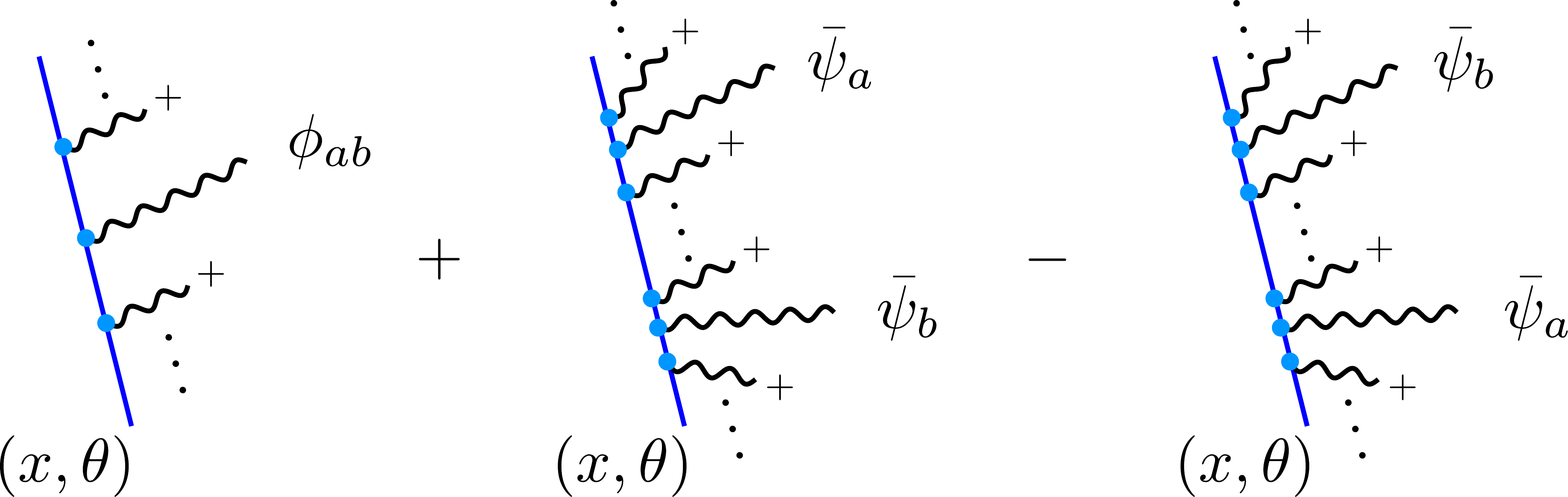}
  \caption{\it The vertex of an operator containing a scalar $\phi_{ab}$ includes all its MHV-preserving splitting terms.}
  \label{fig:Splitting}
\end{figure}

As previously argued,
the twistor-space avatar of any operator $\calOg$ has to contain the terms that allow the elementary fields to split into different ones while preserving the MHV degree. Hence, we propose to complete \eqref{eq:naturalfirstattempt} to
\begin{align}
\label{eq:secondattempt}
&\phi_{ab}(x)=\int \DD\la\,  h_x^{-1}(\la)\frac{\partial^2\calA(\lambda)}{\partial\chi^a\partial\chi^b}h_x(\la)_{\big{|}\theta=0}\notag\\&+\int \frac{\DD\la\DD\la'}{\abra{\la}{\la'}} h_x^{-1}(\la)\frac{\partial\calA(\lambda)}{\partial\chi^a}U_x(\la,\la')\frac{\partial\calA(\lambda')}{\partial{\chi'}^b}h_x(\la')_{\big{|}\theta=0}\notag\\&-(a\leftrightarrow b)\,.
\end{align}
It is depicted in figure \ref{fig:Splitting}. An immediate observation is that in \eqref{eq:secondattempt} the $\chi$-derivatives are now distributed supersymmetrically, unlike in \eqref{eq:naturalfirstattempt}. 

Our proposal for the correction of the expression \eqref{eq:operator} is now obtained by squaring \eqref{eq:secondattempt} and taking the trace. Since it has nine terms, we refrain from writing it out. Using \eqref{eq:secondattempt} and the methods of section \ref{sec:firstattempt} leads to the correct result \eqref{eq:formactorpsisquared}. We can do even better and straightforwardly derive the MHV super form factor of $\calOs$:
\begin{multline}
\label{eq:calFsusy}
\calF_{\calOs}(1,\dots,n;\fq)\\=\frac{\delta^4(\fq-\sum_{k=1}^n\fp_k)\prod_{c=a,b}\left(\sum_{i<j}\abra{i}{j}\eta_{ic}\eta_{jc}\right)}{\abra{1}{2}\abra{2}{3}\cdots \abra{n}{1}}\,.
\end{multline}
This agrees with the results of \cite{Brandhuber:2011tv} and  with \eqref{eq:formfactor2phi} and \eqref{eq:formactorpsisquared} when the respective  components are specified.

Looking at \eqref{eq:secondattempt} and recalling the product rule, one is tempted to try to generate \eqref{eq:secondattempt} by some kind of double derivative. For this, we must replace the operator at point $x$ by a polygonal light-like Wilson loop \cite{Bullimore:2011ni} with an appropriate number of edges $x_1,\ldots, x_n$:
\begin{equation}
\label{eq: polygonal Wilson loop}
 \WWW= \Tr\big[ U_{x_1}(\calZ_1,\calZ_2)
\cdots U_{x_n}(\calZ_n,\calZ_{1}) \big]\,,
\end{equation}
where the $\calZ_i$ are the twistors at the intersection of the line $x_{i-1}$ and $x_i$.
We can then act on $\WWW$ with  four $\theta$-derivatives and finally shrink the Wilson loop back to a point, recovering our expression for $\calOs$. We describe this procedure in full detail in a forthcoming publication \cite{Paper2}, where we also derive the analogues of \eqref{eq:secondattempt} for the rest of the field content of $\mathcal{N}=4$ SYM.

\section{Summary and Outlook}

In this letter, we described how to incorporate composite operators into the twistor-space formulation of $\calN=4$ SYM. Just as translating the action to twistor space shuffles the interaction terms into infinitely many vertices, so does the translation of the composite operators require the repackaging of infinitely many operator vertices. Form factors provide the ideal testing ground for our construction as they are the simplest quantities that contain composite operators. 
Thus, we simultaneously initiated the study of form factors in $\calN=4$ SYM using the position twistor-space framework.

In a forthcoming publication \cite{Paper2}, we use the Wilson loop that we hinted at in section \ref{sec:ourproposal} to derive the tree-level MHV super form factors of all composite operators. We extend the framework to N$^k$MHV form factors and correlation functions in a further publication \cite{Paper3}.

\section*{Acknowledgments}
\label{sec:acknowledgments}
We are thankful to Tim Adamo, Simon Caron-Huot and especially Lionel Mason for insightful comments and  discussions. 
This research is supported in part by the SFB 647 \emph{``Raum-Zeit-Materie. Analytische und Geometrische Strukturen''}. 
L.K. would like to thank the CRST at Queen Mary University of London and especially Gabriele Travaglini for hospitality during a crucial stage of preparation of this work.
V.M. would like to thank the Simons Summer Workshop 2015, where part of this work was performed. 
M.W. was supported  in  part  by  DFF-FNU  through grant number DFF-4002-00037.


\bibliographystyle{JHEP}
\bibliography{biblio}

\end{document}